\documentclass [preprint]{aastex}
\newcommand{\Lya}{Ly$\alpha$}
\newcommand{\MgII}{\ion{Mg}{2}}
\newcommand{\MgI}{\ion{Mg}{1}}
\newcommand{\CaI}{\ion{Ca}{1}}
\newcommand{\CaII}{\ion{Ca}{2}}
\newcommand{\CIII}{\ion{C}{3}]}
\newcommand{\CIV}{\ion{C}{4}}
\newcommand{\FeI}{\ion{Fe}{1}}
\newcommand{\FeII}{\ion{Fe}{2}}
\def\roma#1{\ifmmode{#1}\else{$#1$}\fi} 

\def\kmsmpc{\roma{\rm\,km\,s^{-1}\,Mpc^{-1}}}                       
\newcommand\Ho{\roma{\,\rm H_{\circ}}}                          
\newcommand\qo{\roma{\,\rm q_{\circ}}}                          

\slugcomment{{\em Astronomical Journal, in Press}} 

\begin{document}

\title {{\bf A Close-Separation Double Quasar Lensed by a Gas-Rich
Galaxy} \footnote{Based on observations with the NASA/ESA Hubble Space
Telescope, obtained at the Space Telescope Science Institute, which is
operated by the Association of Universities for Research in Astronomy,
Inc. under NASA contract No. NAS5-26555.}}

\author{Michael~D.~Gregg\altaffilmark{2,3},
Lutz~Wisotzki\altaffilmark{4,5},
Robert~H.~Becker\altaffilmark{2,3},
Jos\'e~Maza\altaffilmark{6},
Paul~L.~Schechter\altaffilmark{5,7},
Richard~L.~White\altaffilmark{8},
Michael~S.~Brotherton\altaffilmark{9}, 
and
Joshua~N.~Winn\altaffilmark{5,7}
}

\altaffiltext{2}{Physics Dept., University of California, Davis, CA
95616; gregg,bob@igpp.llnl.gov}
\altaffiltext{3}{Institute for Geophysics and Planetary Physics,
Lawrence Livermore National Laboratory}
\altaffiltext{4}{Hamburger Sternwarte, Germany; lwisotzki@hs.uni-hamburg.de}
\altaffiltext{5}{Massachusetts Institute of Technology;
schech@achernar.mit.edu, jnwinn@mit.edu}
\altaffiltext{6}{Universidad de Chile; jose@das.uchile.cl}
\altaffiltext{7}{Visiting Astronomer, Cerro Tololo Inter-American
Observatory, National Optical Astronomy Observatories}
\altaffiltext{8}{Space Telescope Science Institute; rlw@stsci.edu}
\altaffiltext{9}{Kitt Peak National Observatory; mbrother@noao.edu}

\begin{abstract}

In the course of a Cycle~8 snapshot imaging survey with STIS, we have
discovered that the z=1.565 quasar HE~0512$-$3329 is a double with
image separation 0\farcs644, differing in brightness by only
0.4~magnitudes.  This system is almost certainly gravitationally
lensed.  Although separate spectra for the two images have not yet
been obtained, the possibility that either component is a Galactic
star is ruled out by a high signal-to-noise composite ground-based
spectrum and separate photometry for the two components: the spectrum
shows no trace of any zero redshift stellar absorption features
belonging to a star with the temperature indicated by the broad band
photometry.  The optical spectrum shows strong absorption features of
\MgII, \MgI, \FeII, \FeI, and \CaI, all at an identical intervening
redshift of z=0.9313, probably due to the lensing object.  The
strength of \MgII\ and the presence of the other low-ionization
absorption features is strong evidence for a damped \Lya\ system,
likely the disk of a spiral galaxy.  Point spread function fitting to
remove the two quasar components from the STIS image leads to a
tentative detection of a third object which may be the nucleus of the
lensing galaxy.  The brighter component is significantly redder than
the fainter, due to either differential extinction or microlensing.

\end{abstract}


\section {Introduction}

The study of gravitationally lensed quasars has become a powerful tool
for addressing a number of astrophysical questions.  In particular,
concentrating on studying the lensing objects themselves provides a
sample of distant galaxies selected by mass rather than by light
(Kochanek et al.\ 1999).  Because the component separations scale with
the square root of the mass of the lens, sampling the low end of the
lens mass function becomes difficult from the ground, particularly in
the optical/IR, for separations $\lesssim 1\arcsec$.  This
observational bias leads to a preponderance of massive spheroidal
galaxies in the present sample of lenses and at least partly accounts
for the relative lack of known close-separation lenses, which are
predicted to exist by theoretical models of the lensing phenomenon
(e.g.\ Maoz \& Rix 1993; Rix et al.\ 1994; Jain et al.\ 1999).  There
are currently only seven systems with separations $\leq 0\farcs9$ out
of 43 confirmed lensed quasars listed by Kochanek et al.\ (1998).

We are now well into a Cycle~8 snapshot survey of up to 300 targets,
aimed specifically at finding close-separation lensed quasars using
the imaging capabilities of STIS (Kimble et al.\ 1997; Woodgate et
al.\ 1998) on board the Hubble Space Telescope.  The probability that
a quasar is lensed increases with redshift and apparent magnitude
(Turner, Ostriker, \& Gott 1984); the snapshot survey targets bright,
high redshifts quasars selected using estimates for the probability of
lensing by Kochanek (1998).  The results of the full snapshot survey
will appear in time (Gregg et al.\ 2000, in prep.); here we report the
discovery of a close-separation gravitationally lensed quasar from
among the first 80 snapshot targets.

\section {Observations}

\subsection {Discovery}

The quasar HE~0512$-$3329 was originally identified in the Hamburg/ESO
survey for bright QSO's (Wisotzki et al.\ 1996).  With $B$ = 17.0 and z
= 1.569 (Reimers, K\"{o}hler, \& Wisotzki 1996), HE~0512$-$3329 had an
a priori probability of $\sim 1.3\%$ of being lensed, fairly typical
for the targets in our snapshot survey.  The STIS snapshot sequence,
obtained on 1999 August 26, consists of $3 \times 40$s CR-split
exposures in the clear 50CCD (CL) aperture and one additional 80s
CR-split exposure in the longpass F28$\times$50LP (LP) filter.  The effective
wavelength and full width half maximum of the CL band are 6167.6\AA\
and 4410\AA\ and for the LP band are 7333\AA\ and 2721\AA.  The STIS
images reveal two point sources with a separation of 0\farcs644.  The
difference in brightness between the two components~A and B is $\Delta
CL = 0.35$ and $\Delta LP = 0.49$; this small difference is
characteristic of the more highly magnified lensed systems, which our
selection technique is designed to favor.

The lensing hypothesis was strengthened by examining the discovery
spectrum of HE~0512$-$3329 which shows a rather typical quasar energy
distribution having emission lines of \CIV~1549 and \CIII~1909 (see
Figure~1 of Reimers et al.\ 1996).  If one of the two components were
a garden variety Galactic star, the strongest stellar absorption lines
would be easily identifiable (see below, \S 2.3).  A binary quasar is
a possible alternative explanation (Kochanek, Falco, \& Mu\~{n}oz
1999), however, the discovery spectrum exhibits a strong absorption
feature consistent with \MgII\ at an intervening redshift of 0.93, and
a few weaker absorption lines of \FeII\ at the same redshift.  These
low-ionization absorption features suggested that the duplicity is due
to a lensing object at this redshift.

\subsection {Follow-up Spectroscopy at Keck Observatory}

In early 2000 January, we obtained a 9\AA\ resolution spectrum of
HE~0512$-$3329 using the Low Resolution Imaging Spectrograph (LRIS,
Oke et al.\ 1995) at Keck Observatory.  The slit was oriented at the
position angle of the two quasar images on the sky, 17\arcdeg.  The
seeing was 1\arcsec, insufficient to resolve the components.  From
this 300s exposure (Figure~1), we obtain a redshift of $1.565 \pm
0.001$ based on Gaussian fits to the \CIV\ and \CIII\ emission peaks.
The S/N of this new spectrum is 50 to 100 over most of its wavelength
range and confirms the presence of the strong intervening absorption,
clearly resolving the \MgII~2796.4, 2803.5 doublet and detecting the
associated \MgI~2853 line.  Also seen is a rich absorption system of
\FeII~2260.8, 2344.2, 2374.5, 2382.8, 2586.7, and 2600.2, and
\CaI~4227.9 belonging to the same intervening system; \CaII~3933 and
3969 fall in the atmospheric A band.  The mean of the absorption
redshifts is z = $0.9313 \pm 0.0005$.  The profile of the \MgII\
emission line is asymmetric, which can be attributed to absorption by
\FeI~3721.0 or intervening \MgII\ local to the quasar.  There is
another weak intervening \MgII\ absorption feature at z=1.1346.

\begin{figure}[p]
\plotone{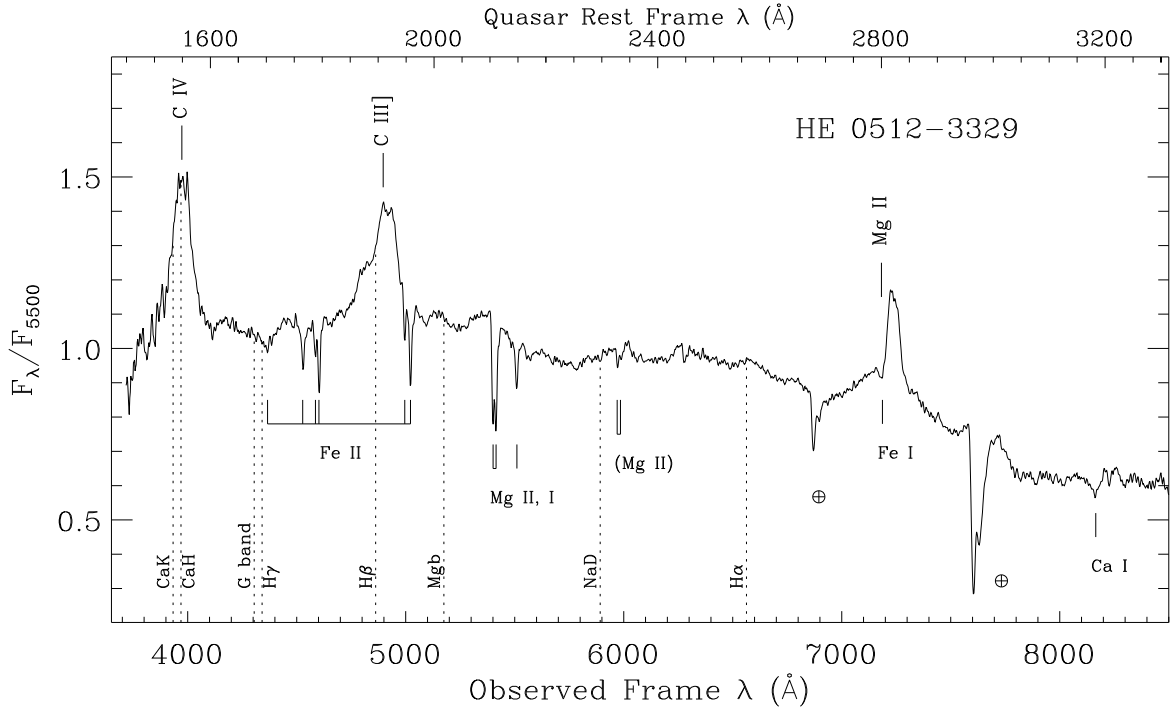}
\caption{Keck LRIS 9\AA\ resolution spectrum of HE~0512$-$3329
obtained in January, 2000.  The upper abscissa scale is the rest frame
of the quasar with a redshift of 1.565; prominent emission line
features are marked above the spectrum.  Lines associate with the
intervening system at z=0.9313 are identified below the spectrum.  The
absorption lines are probably due to the lensing galaxy and are
unresolved at the instrumental resolution.  There is a second rather
weak intervening \MgII\ 2800 absorption system at z=1.1346, marked in
parenthesis.  The small difference in brightness of the two components
and the lack of any visible stellar absorption features (dotted lines)
in this high S/N composite spectrum argues strongly against component
B being a foreground star.}
\end{figure}

Component~B is 70\% the brightness of A.  The lack of any discernible
stellar absorption features in the Keck spectrum (Figure~1) argues
strongly against component~B being a foreground star.  The RMS noise
in the spectrum is at the level of 0.15\AA\ equivalent width.  The
strongest features in late type stars have equivalent widths of a few
\AA ngstroms and would be detected easily in the Keck spectrum; for
comparison, the equivalent width of the intervening \MgI\ 2853\AA\
feature is 1.4\AA.  The only possible stellar contaminant is a
completely featureless O-type subdwarf or white dwarf and such stars
are extremely rare.  If this were the case, however, the spectrum of
HE~0512$-$3329 would be much bluer, unless either the QSO or the
putative star has a large amount of intrinsic reddening.

\subsection {Photometry}

We have done point-spread function (PSF) fitting photometry on the
STIS images using IRAF/DAOPHOT.  From observations of an unlensed
quasar in our program, we obtain aperture corrections of $-0.222$ and
$-0.303$ for the CL and LP bands, to go from the fitted 3 pixel radius
to 0\farcs5.  To this we add an additional $-0.1$ magnitudes to correct
to the ``true'' magnitude in an infinite aperture, as is standard
practice with WFPC2 (Holtzman et al.\ 1996).  The resulting calibrated
STIS ``STmagnitudes'' and errors are listed in Table~1.

The $LP$ bandpass is completely contained within the $CL$.  Because
they are similar in shape, the $LP$ flux can be scaled by the relative
throughputs and subtracted from the $CL$, producing an effective
``shortpass'' measurement (Gregg \& Minniti 1997; Gardner et al.\
2000) extending from 5500\AA\ to 2000\AA, with effective wavelength of
4424\AA\ and FWHM of 2569\AA.  The $SP-LP$ difference provides some
wide-band color information (Table~1).

In 1999 December, we obtained VRI photometry of HE~0512$-$3329 using
the Mosaic~II CCD imager at the Blanco 4m telescope at Cerro
Tololo Inter-American Observatory\footnote{Cerro Tololo Inter-American
Observatory, NOAO, is operated by the Association of Universities for
Research in Astronomy, Inc. (AURA), under cooperative agreement with
the National Science Foundation.}.  Although the seeing was $\sim
0\farcs8 - 0\farcs9$ and the two components are not cleanly resolved,
point-spread function (PSF) fitting using IRAF/DAOPHOT successfully
separated them, yielding positions in excellent agreement with the HST
images and photometry consistent with the STIS results.  The separations
obtained in V, R, and I are 0\farcs654, 0\farcs646, and 0\farcs643,
respectively, compared to 0\farcs644 obtained from the centroids of
the components in the STIS CL images.

No photometric standards were taken at CTIO, so we have calibrated the
CTIO photometry using zeropoints determined by convolving the Keck
spectrophotometry with Cousins VRI passbands.  This procedure is
itself calibrated using a model for the spectrum of Vega (Kurucz 1992)
for which we adopt $B = V = R = I = 0.0$.  Slit losses limit the
absolute accuracy, but, fortuitously, the spectroscopy was obtained
when the position angle of HE~0512$-$3329 was only 23\arcdeg\ from the
parallactic angle.  Because the effective slit width was somewhat
greater than the atmospheric dispersion between the red and blue
extremes of the spectrum (Filippenko 1982), the colors obtained from
the composite spectrum are reasonably accurate and can be used to
establish the relative zeropoints of the VRI photometry.  Also, we
have determined a zeropoint transformation between the effective STIS
SP bandpass and Johnson B using the mean quasar spectrum (Brotherton
et al.\ 2000) from the FIRST Bright Quasar Survey (FBQS; White et al.\
2000), redshifted to z=1.565.  The resulting colors of HE~0512$-$3329
are $B-V = 0.68, V-R = 0.46, V-I = 0.82$ for component~A, and $B-V =
0.32, V-R = 0.37, V-I = 0.69$ for component~B (Table~1).  Schlegel,
Finkbeiner, \& Davis (1998) estimate a Galactic extinction of $A_B =
0.104$ for this line of sight; Burstein \& Heiles (1982) give a much
lower value of 0.010.  The numbers in Table~1 have not been corrected
for Galactic extinction.

\begin {deluxetable}{lrrrrrrrrr}
\tabletypesize{\small}
\tablewidth{0pc}
\tablecaption{HE 0512$-$3329}
\tablehead{
\colhead{Object} &
\colhead{R.A. (J2000)} &
\colhead{Dec.\ (J2000)} &
\colhead{$CL$} &
\colhead{$LP$} &
\colhead{$SP$} &
\colhead{$B$} &
\colhead{$V$} &
\colhead{$R$} &
\colhead{$I$}
}
\startdata
A & 05 14 10.7833  & $-33$ 26 22.504 & 17.94 &  18.53  & 19.47 &
18.36 & 17.68 & 17.22 & 16.86  \\
B & 05 14 10.7687  & $-33$ 26 23.121 & 18.28 &  19.03  & 19.46 &
18.35  & 18.03 & 17.66 & 17.34 \\
B-A & -0\farcs183  & 0\farcs618 & 0.35 &  0.49  & -0.01 &
-0.01  & 0.35 & 0.44 & 0.48 \\
errors & 0\farcs003  & 0\farcs003 & 0.02 &  0.02  & 0.02 &
0.04  & 0.02 & 0.02 & 0.02 \\
\enddata
\tablecomments{Photometry is uncorrected for Galactic reddening; errors
are statistical only.}
\end {deluxetable}

\begin {deluxetable}{lccrrr}
\tabletypesize{\small}
\tablewidth{0pc}
\tablecaption{Field Star Difference Photometry}
\tablehead{
\colhead{Object} &
\colhead{R.A. (J2000)} &
\colhead{Dec.\ (J2000)} &
\colhead{$\Delta V_i$} &
\colhead{$\Delta R_i$} &
\colhead{$\Delta I_i$}
}
\startdata
Star 1 $-$ A &  5 14 08.98 & $-33$ 27 08.1  &     -1.04  &   -1.14  &   -1.12 \\   
Star 2 $-$ A &  5 14 11.62 & $-33$ 26 50.1  &      0.56  &    0.77  &    0.95 \\   
Star 3 $-$ A &  5 14 14.83 & $-33$ 27 24.7  &      0.91  &    0.93  &    0.96 \\   
Star 4 $-$ A &  5 14 18.74 & $-33$ 26 03.6  &     -0.38  &   -0.38  &   -0.32 \\   
Star 5 $-$ A &  5 14 19.06 & $-33$ 26 13.6  &      1.33  &    0.99  &    0.72 \\  
Star 6 $-$ A &  5 14 08.63 & $-33$ 24 26.9  &     -0.30  &   -0.17  &   -0.09 \\  
Star 7 $-$ A &  5 14 08.51 & $-33$ 24 48.4  &      0.81  &    0.78  &    0.77 \\  
Star 8 $-$ A &  5 13 57.75 & $-33$ 26 24.3  &      1.25  &    0.88  &    0.54 \\  
Star 9 $-$ A &  5 14 05.42 & $-33$ 29 15.7  &     -0.58  &   -0.42  &   -0.29 \\  
\enddata
\tablecomments{Numbers are instrumental magnitude differences $m_{star} - m_A$
in the Cousins system; no color terms have been applied.}
\end {deluxetable}

The broad band colors of the two components clinch the case for
HE~0512$-$3329 being a lensed quasar.  By comparison with the Bruzual
et al.\ stellar library, the $B-V$ color of component~B indicates a
spectral type of F0, yet V-R and V-I are consistent with a much cooler
object, about F9/G0.  Our simulations show that any star in this
spectral range with the relative brightness of component~B would
contribute easily detectable absorption features, at $\sim 10\sigma$
level or greater, to the composite spectrum at the indicated locations
in Figure~1; \CaII~H and K and the Balmer lines would be particularly
conspicuous.  The broad band colors of component~B are, in fact, more
consistent with a slightly reddened quasar than a star.

For future reference for monitoring variability of the lens
components, we list in Table~2 the instrumental magnitude differences,
$m_i - m_{\rm A}$, for nine field stars with $V \approx 16$ to 18 in
the vicinity of HE~0512$-$3329.  The astrometry has been derived from
the digitized sky survey and has an offset of $\Delta R.A. =
+0\farcs55$ and $\Delta Dec.\ = +0\farcs77$ relative to the STIS
images, but these positions are sufficient to unambiguously identify 
the comparison stars.

\section{The Nature of the Lensing Object}

The presence of the strong intervening \MgII\ absorption and the many
associated low ionization lines are evidence that the lensing object
contains a damped \Lya\ absorption (DLA) system (Boisse et al.\ 1998).
A DLA system at a redshift $< 1$ is most likely to be the
hydrogen-rich disk of a spiral galaxy.  Dust in the galaxy may produce
differential reddening in the two components of HE~0512$-$3329.

\subsection{Possible Detection of a Third Object}

To explore for the lensing galaxy and possible additional quasar
images, the STIS $CL$ images were combined using the {\sc DRIZZLE}
package (Fruchter \& Hook 1998) in IRAF/STSDAS.  A sampling rate of
0.5 times the original image scale and a {\sc PIXFRAC} value of 0.6
were used.  The subpixel image shifts were determined using the IRAF
task {\sc XREGISTER}.  The final combined $CL$ image is shown in the
left panel of Figure~2.  The distance between centroids of the two
images is 0\farcs644; at the probable lens redshift of 0.9313, this
separation is only $\sim 4$ kpc, adopting \Ho = 70 \kmsmpc and \qo =
0.5.  The mass associated with an Einstein ring of this scale is $\sim
3 \times 10^{10}$ M$_\odot$.

\begin{figure}[t]
\plotone{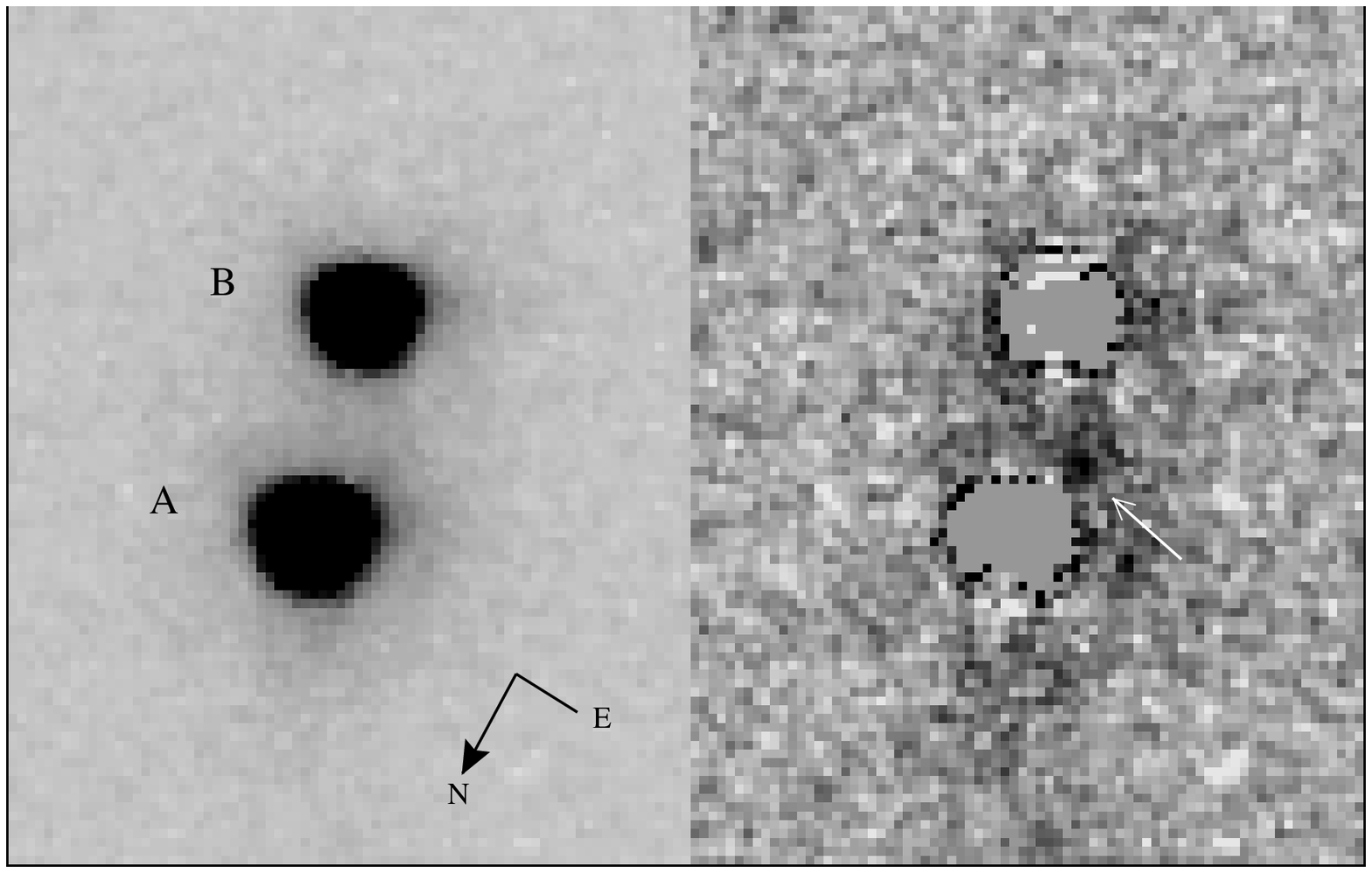}
\caption{{\bf Left:} STIS 50CCD ($CL$) image of HE~0512$-$3329.
Component~A is 0.35 magnitudes brighter than B in this passband.  The
separation is 0\farcs644.  The orientation is given by the compass
points; the North arrow is 0\farcs25 long. \newline 
{\bf Right:} Residuals
after PSF removal.  The object immediately above and to the right of
A, indicated by the white arrow, may be the nucleus of the lensing
galaxy.  There is also excess low surface brightness flux around each
component.}
\end{figure}

The PSF removal was done using the {\sc SCLEAN} task in IRAF/STSDAS.
For the PSF itself, we used the theoretical STIS PSF from the {\bf
Tiny Tim} package and also the STIS PSF generated by the Hubble Deep
Field South project (Gardner et al.\ 2000).  They yield very similar
results.  The residual image is shown in the right hand panel of
Figure~2.  There is an excess of counts just above and to the right of
component~A (white arrow in Figure~2).  The RMS in the
background-subtracted image is $\sim \pm 1.3$ counts while the peak in
the excess region is 8.7.  The total flux is $\sim 5.9$ magnitudes
fainter than component~A, giving it $CL = 23.8$.  Its FWHM is roughly
twice that of a point source, consistent with being nonstellar.  We
tentatively identify this object as the nucleus of the lensing galaxy.
For our adopted cosmology, this object has $M_V \approx -19$, roughly
the nucleus of a roughly $L^*$ galaxy with a bulge-to-disk ratio of
$\sim 3$ and, for the above quoted Einstein ring mass, a M/L ratio of
$\sim 20$.  At the intervening redshift of 0.9313, the the two lines
of sight to the quasar pass 1.6 (A) and 2.7 (B) kpc from the position
of the third object.  

There is excess light of lower surface brightness between the two
quasar images and immediately below component~A as well as to the
right of component~B in Figure~2.  Although the third object is not
detected with confidence in the LP image, the low surface brightness
light distribution is qualitatively reproduced in the redder passband.
No such excess light is seen when the same analysis is applied to an
image of an unlensed quasar from our snapshot program.  Deeper images
are needed to confirm the reality of this low surface brightness fuzz
and, if real, determine whether it is due to the lens, the host
galaxy, or another object.

\subsection{Reddening Analysis}

The STIS and ground-based photometry are consistent in showing that
component~A is redder than B.  Going from red to blue, the magnitude
difference between the two images decreases, becoming equal within the
errors in the concocted STIS $SP$ and transformed $B$ bands
(Figure~3).  Observed in the ultraviolet, component~B will be the
brighter.  This trend can be attributed to differential reddening,
with extinction along the line of sight to component~A being greater.
Color differences between the quasar images can also be arise from
microlensing by stars in the lensing galaxy, producing differential
magnification of the quasar continuum (Wambsganss \& Paczy\'nski 1991)
along the two lines of sight.  This effect has been observed in at
least one quasar, HE~1104$-$1805 (Wisotzki et al.\ 1993).  The
following reddening analysis is valid only if microlensing is
negligible in HE~0512$-$3329.

\begin{figure}[t]
\plotone{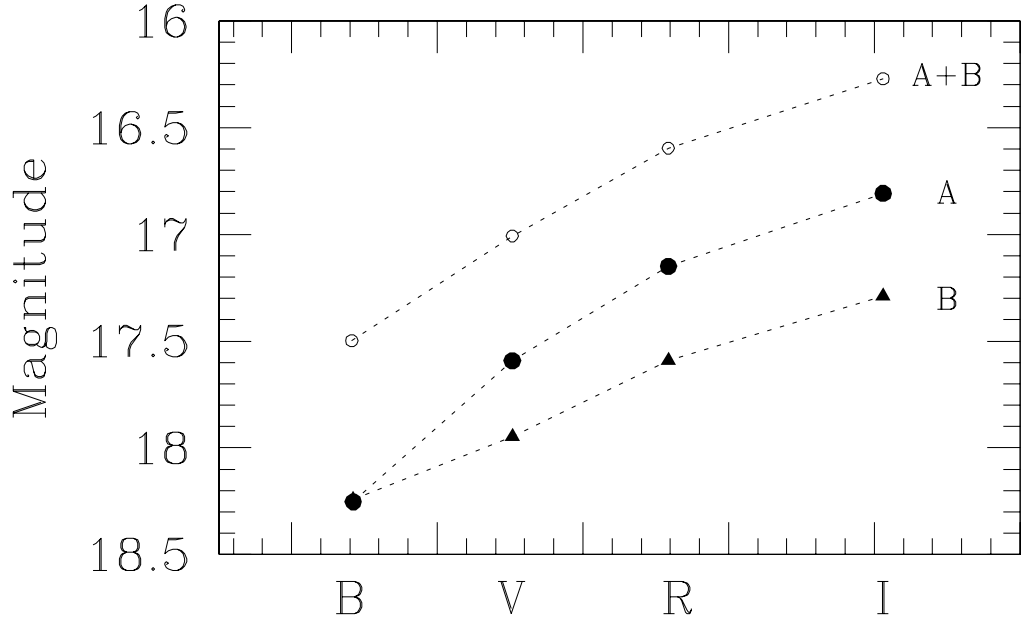}
\caption{Trend of the broad band magnitudes of HE~0512$-$3329 for
components A, B, and the composite A+B.  The VRI photometry is from
the ground-based CTIO imaging while the B magnitudes are derived from
the difference between the STIS CL and LP bands; both have been
zeropointed using the Keck spectrophotometry in Figure~1 (see text for
details).  The brightness of the two images converges from red to
blue.  This can be attributed to differential extinction or
microlensing, possibly both.  These data have been corrected from
Table~1 for Galactic absorption of $A_B = 0.104$}
\end{figure}

To quantify the amount of differential reddening, we have fitted an
extinction model to the photometry results, following the procedure of
Falco et al.\ (1999).  In this approach, it is assumed that the quasar
is not variable, that the lensing magnification is not wavelength
dependent, and that the extinction law does not vary with position in
the lensing galaxy and is well-approximated by a typical Galactic
extinction curve with $R_V = A_V/E(B-V) = 3.1$.  Correcting a sign
error in equation~3 of Falco et al., we have
\begin{equation}
\chi^2 = \sum_{j=1}^{N_{\lambda}}\sum_{i=1}^{N_c} \frac{\{m_i(\lambda_j) - m_{\circ}(\lambda_j)
+ 2.5\log(M_i) - E_i~R[\lambda_j/(1+z)]\}^2}{\sigma_{ij}^{2}}
\end{equation}
where $m_i$ are the observed magnitudes of each of the $N_c$
components in the $N_{\lambda}$ photometric bands with effective
wavelength $\lambda_j$, $m_{\circ}$ is the unlensed magnitude of the
quasar, $M_i$ is magnification of each component, $E_i$ is the
extinction of each component, z is the redshift of the lens, and
$\sigma_{ij}$ are the photometric errors.  The summations are over the
4 bandpasses, $B, V, R,$ and $I$, and two components, A and B.

Relative magnifications and extinctions can be found by minimizing
$\chi^2$ while holding one magnification fixed at unity and one
extinction at 0.  As component~B is bluer and fainter, we fix its
parameters at these values.  The extinction law has been parametrized
using the equations of Cardelli, Clayton, \& Mathis (1989).  For this
analysis, we first corrected the photometry listed in Table~1 for
Galactic extinction of $A_B = 0.104$ from Schlegel et al.\ (1998).

The fit for the relative extinction results in the estimate of $A_V =
0.34$ for component~A, in excess of the extinction at component~B; the
unextincted, wavelength independent relative magnification of A is
2.45 times that of B.  The $\chi^2$ of this fit is 0.66.  For
comparison, a fit with both extinctions held to zero yields a relative
magnification of 1.35, roughly consistent with the brightness
difference between the two components in $V$ or $R$.  The $\chi^2$ for
this fit is 67, as might be expected given the varying magnitude
difference between the two components as a function of wavelength,
which renders an achromatic magnification model a poor explanation of
the brightness variation with wavelength.

The separate extinctions to each component of HE~0512$-$3326 can be
estimated by assuming that the unlensed quasar spectrum has typical
colors.  After correcting for Galactic reddening using the Schlegel et
al.\ (1998) value, the difference in $B-V$ between the composite
spectrum of HE~0512$-$3326 (Figure~1) and the FBQS mean spectrum is
0.31.  With $R_V = 3.1$, this is equivalent to $A_V = 0.97$.  Knowing
the $V$ magnitudes and relative extinction, the separate extinctions
can be computed as $A_V^{\rm A} = 1.10$ and $A_V^{\rm B} = 0.76$,
excluding any grey component.  Given the multitude of assumptions and
the bootstrapping from the spectrophotometry, these numbers must be
considered provisional, but they do suggest that the extinction to
each component is comparable and that both lines of sight intercept
the same or similar absorption systems.  Spectroscopy of the two
components separately would allow a detailed study of the extinction
curve in the disk of the lensing galaxy and would also determine
whether microlensing could be contributing to the pattern of color
differences.

\section {Conclusion}

The presently available data leave little doubt that HE~0512$-$3329 is
gravitationally lensed.  The spectroscopic evidence strongly suggests
that the lens is a spiral galaxy.  Spatially resolved spectroscopy of
the two images of HE~0512$-$3329 is needed to confirm its nature;
however, the presence of strong low-ionization lines in the composite
spectrum indicates that at least one of the lines of sight is sure to
pass through a damped Ly$\alpha$ system in the disk of the lens.
Ultraviolet spectroscopy of the A and B components can further be used
to derive the extinction curve in the disk of the lens as well as
abundances of heavy elements.  This lensed quasar has been found among
the first 80 targets of an HST Cycle~8 snapshot program designed to
search for such small separation systems.  The program was renewed for
up to 300 additional snapshots in Cycle~9.  If close-separation
targets are found with this frequency for the duration of the survey,
the lensing statistics for small separation systems will be boosted by
a significant factor.

\acknowledgments

Mark Lacy is thanked for helpful discussions.  The referee is credited
with constructive comments which improved this paper.  We are grateful
to Sune Toft for calling our attention to an error in our original
calculation of the differential reddening.  Support for this work was
provided by NASA through grant number GO-8202 from the Space Telescope
Science Institute, which is operated by AURA, Inc., under NASA
contract NAS5-26555.  We also acknowledge support from NSF grant
AST-98-02791.  This work was performed under the auspices of the
U.S. Department of Energy by University of California Lawrence
Livermore National Laboratory under contract No.~W-7405-Eng-48.
J.N.W. thanks the Fannie and John Hertz Foundation for financial
support.

\end{document}